\documentclass[article,preprint,groupedaddress]{revtex4}
\usepackage{epsfig}
\usepackage{graphicx}

\begin{document}
\title{Gravitation, Thermodynamics, and the Bound on Viscosity}
\author{Shahar Hod}
\address{The Ruppin Academic Center, Emeq Hefer 40250, Israel}
\address{ }
\address{The Hadassah Institute, Jerusalem 91010, Israel}
\date{\today}
{\it Essay written for the Gravity Research Foundation 2009 Awards
for Essays on Gravitation}

\begin{abstract}
\ \ \ The anti-de Sitter/conformal field theory (AdS/CFT)
correspondence implies that small perturbations of a black hole
correspond to small deviations from thermodynamic equilibrium in a
dual field theory. For gauge theories with an Einstein gravity dual,
the AdS/CFT correspondence predicts a universal value for the ratio
of the shear viscosity to the entropy density, $\eta/s=1/4\pi$. It
was conjectured recently that all fluids conform to the lower bound
$\eta/s \geq 1/4\pi$. This {\it conjectured} bound has been the
focus of much recent attention. However, despite the flurry of
research in this field we still lack a proof for the general
validity of the bound. In this essay we show that this mysterious
bound is actually a direct outcome of the interplay between gravity,
quantum theory, and thermodynamics.
\newline
\newline
Email: shaharhod@gmail.com
\end{abstract}
\bigskip
\maketitle


The anti-de Sitter/conformal field theory (AdS/CFT) correspondence
\cite{Mald} has yielded remarkable insights into the dynamics of
strongly coupled gauge theories. According to this duality,
asymptotically AdS background spacetimes with event horizons are
interpreted as thermal states in dual field theories. This implies
that small perturbations of a black hole or a black brane background
correspond to small deviations from thermodynamic equilibrium in a
dual field theory. One robust prediction of the AdS/CFT duality is a
universally small ratio of the shear viscosity to the entropy
density \cite{Poli},
\begin{equation}\label{Eq1}
{\eta \over s}={1 \over{4\pi}}\  ,
\end{equation}
for all gauge theories with an Einstein gravity dual in the limit of
large 't Hooft coupling. (We use natural units for which
$G=c=\hbar=k_B=1$.)

It was suggested \cite{Poli} that (\ref{Eq1}) acts as a universal
lower bound [the celebrated Kovtun-Son-Starinets (KSS) bound] on the
ratio of the shear viscosity to the entropy density of general,
possibly nonrelativistic, fluids. Currently this bound is considered
a {\it conjecture} well supported for a certain class of field
theories \cite{BekFoux}. So far, all known materials satisfy the
bound for the range of temperatures and pressures examined in the
laboratory. The system coming closest to the bound is the
quark-gluon plasma created at the BNL Relativistic Heavy Ion
Collider (RHIC) \cite{Tean}. In fact, it was the challenge presented
by the quark-gluon plasma which motivated the activity leading to
the formulation of the KSS bound (1).

Where does the KSS bound on the ratio $\eta/s$ come from ? It is not
clear how to obtain such a bound directly from microscopic physics
\cite{BekFoux}. Inspection of the Green-Kubo formula \cite{Land}
which relates the viscosity of a fluid to its fluctuations shows no
apparent connection of the viscosity and the entropy density. Such
microscopic consideration affords no special status to the ratio
$\eta/s$.

Where should we look for the physical mechanism which bounds the
ratio $\eta/s$ of viscosity to entropy density ? It is well known
that the viscosity coefficient $\eta$ characterizes the intrinsic
ability of a perturbed fluid to relax towards equilibrium
\cite{Dani} [see Eq. (\ref{Eq3}) below]. The response of a medium to
mechanical excitations is characterized by two types of normal
modes, corresponding to whether the momentum density fluctuations
are transverse or longitudinal to the fluid flow. Transverse
fluctuations lead to the shear mode, whereas longitudinal momentum
fluctuations lead to the sound mode. These perturbation modes are
characterized by distinct dispersion relations which describe the
poles positions of the corresponding retarded Green functions
\cite{Poli}.

Let us examine the behavior of the shear mode for fluids with zero
chemical potential. The Euler identity reads $\epsilon+P=Ts$, where
$\epsilon$ is the energy density, $P$ is the pressure, $T$ is the
temperature, and $s$ is the entropy density of the fluid. The
dispersion relation for a shear wave with frequency $\omega$ and
wave vector $k\equiv 2\pi/\lambda$ is given by \cite{Baier,Nats}:
\begin{equation}\label{Eq3}
\omega(k)_{\text{shear}}=-i{{\eta}\over{Ts}}k^2+O\Big({{\eta^3k^4}\over{s^3T^3}}\Big)\
,
\end{equation}
where $\eta$ is the shear viscosity coefficient of the standard
first-order hydrodynamics. The correction term becomes small in the
$\eta/s\ll 1$ limit, the case of most interest here.

The imaginary part of the dispersion relation entails a damping of
the perturbation mode. Its magnitude therefore quantifies the
intrinsic ability of a fluid to dissipate perturbations and to
approach thermal equilibrium.

It is important to realize that hydrodynamics is actually an
effective theory. In the most common applications of hydrodynamics
the underlying microscopic theory is a kinetic theory. In these
cases the microscopic scale which limits the validity of the
effective hydrodynamic description is the mean free path
$l_{\text{mfp}}$ \cite{Baier}. In some other cases the underlying
microscopic theory is a quantum field theory, which might not
necessarily admit a kinetic description. In these cases, the role of
the parameter $l_{\text{mfp}}$ is played by some typical microscopic
scale like the inverse temperature: $\l_{\text{mfp}}\sim T^{-1}$.
One therefore expects to find a breakdown of the effective
hydrodynamic description at spatial and temporal scales of the order
of \cite{Baier}
\begin{equation}\label{Eq4b}
l\sim \tau\sim T^{-1}\  .
\end{equation}
Below we shall make this statement more accurate.

At this point, it is worth emphasizing that the conjectured KSS
bound is based on holographic calculations of the shear viscosity
for strongly coupled quantum field theories with gravity duals
\cite{Mald,Poli}. These holographic arguments serve to connect
quantum field theory with gravity. This fact indicates that a
derivation of the KSS bound may require use of the elusive theory of
quantum gravity. This may seem as bad news for our aspirations to
prove a KSS-like bound. But one need not loose heart-- there is
general agreement that black hole entropy reflects some aspect of
the quantum theory of gravity \cite{BekFoux}.

The realization that a black hole is endowed with well-defined
entropy $S_{BH}=A/4$, where $A$ is the surface area of the black
hole \cite{Bek1}, has lead to the formulation of the generalized
second law (GSL) of thermodynamics. The GSL is a unique law of
physics that bridges thermodynamics and gravity \cite{Bek1,BekFoux}.
It asserts that in any interaction of a black hole with an ordinary
matter, the sum of the entropies (matter+hole) never decreases. One
of the most remarkable predictions of the GSL is the existence of a
universal entropy bound \cite{Bek4}. According to this universal
bound, the entropy contained in a given volume should be bounded
from above:
\begin{equation}\label{Eq5}
S\leq 2\pi RE\  ,
\end{equation}
where $R$ is the effective radius of the system and $E$ is its total
energy.

Furthermore, the generalized second law allows one to derive in a
simple way two important new quantum bounds:
\begin{itemize}
\item{The universal relaxation bound \cite{Hod1,Gruz}. This bound asserts that the
relaxation time of a perturbed thermodynamic system is bounded from
below by
\begin{equation}\label{Eq6}
\tau\geq 1/\pi T\  ,
\end{equation}
where $T$ is the temperature of the system. This bound can be
regarded as a quantitative formulation of the third law of
thermodynamics. One can also write this bound as $\Im\omega\leq\pi
T$, where $\omega$ is the quasinormal frequency of the perturbation
mode.}
\item{A closely related conclusion is that thermodynamics can not be
defined on arbitrarily small length scales. The minimal length scale
(radius) $\ell$ for which a consistent thermodynamic description is
available is given by $\ell_{min}=1/2\pi T$ \cite{Gruz,BekFoux}.}
\end{itemize}

The longest wavelength which can fit into a space region of
effective radius $\ell$ is $\lambda_{max}=2\pi\ell$. Thus, the GSL
predicts that an effective hydrodynamic description is limited to
perturbation modes with wavelengths larger than
$2\pi\ell_{min}=T^{-1}$. We note that this exact limit is in accord
with the heuristic arguments which lead to the approximate relation
(\ref{Eq4b}). The breakdown of the effective hydrodynamic
description for perturbation modes with wavenumbers $k$ larger than
$2\pi T$ should manifest itself in the hydrodynamic dispersion
relation (\ref{Eq3}). Namely, one should expect to find a violation
of the universal relaxation bound (\ref{Eq6}) by short wavelength
perturbations with $k>2\pi T$.

A lower bound on the ratio $\eta/s$ can be inferred by substituting
$k=2\pi T$ in the shear dispersion relation (\ref{Eq3}) and
requiring that $\Im\omega\geq \pi T$ for this limiting value of the
wavenumber. As discussed above, the GSL predicts that the effective
hydrodynamic description breaks down for short wavelength
perturbations with $k\geq 2\pi T$. This should be reflected in the
hydrodynamic shear dispersion relation in the form of a violation of
the universal relaxation bound (\ref{Eq6}). Explicitly, these
wavenumbers should be characterized by $\Im\omega\geq\pi T$. This
physical condition implies the inequality
\begin{equation}\label{Eq7}
{\eta \over s}\geq {1 \over{4\pi}}\ .
\end{equation}
We have therefore proved the previously conjectured
viscosity-entropy bound.

{\it Summary.---} The conjectured bound on viscosity has been the
focus of much recent attention. However, despite the flurry of
research in this field the origin of the bound has remained somewhat
unclear. In this essay we showed that the GSL, a law whose very
meaning stems from gravitation, may shed much light on the origins
of this mysterious bound. In particular, it was shown that the
viscosity bound is actually a direct outcome of the interplay
between gravity, quantum theory, and thermodynamics.

\bigskip
\noindent
{\bf ACKNOWLEDGMENTS}
\bigskip

This research is supported by the Meltzer Science Foundation. I
thank D. T. Son and A. O. Starinets for helpful correspondence. I
also thank Yael Oren for stimulated discussions.


\end{document}